\documentclass[showpacs,twocolumn,showkeys,amsmath,amssymb,pra]{revtex4} 

\pdfoutput=1
\usepackage{graphicx}   
\usepackage{dcolumn}    
\usepackage{bm}         
\begin{document} 
 
\title{Reproducible mesoscopic superpositions of Bose-Einstein condensates and mean-field chaos} 

\author{Bettina Gertjerenken}
\email{gertjerenken@theorie.physik.uni-oldenburg.de}
\author{Stephan Arlinghaus} 
\author{Niklas Teichmann}
\author{Christoph Weiss}

\affiliation{Institut f\"ur Physik, Carl von Ossietzky Universit\"at,
                D-26111 Oldenburg, Germany
}

                  
\date{\today}
 
\begin{abstract}

In a parameter regime for which the mean-field (Gross-Pitaevskii) dynamics becomes
chaotic, 
mesoscopic quantum superpositions in phase space can occur in a double-well potential which is
shaken periodically. For
experimentally realistic initial states like the ground state 
of some 100 atoms, the emergence of mesoscopic quantum superpositions in phase space is investigated
numerically. It is shown to be reproducible even if the initial conditions slightly
change. While the final state is not a perfect superposition of two distinct phase-states, the
superposition is reached
an order of magnitude faster than in the case of the collapse and revival phenomenon. Furthermore, a generator of entanglement is identified.
\end{abstract} 


\maketitle


\section{Introduction}

Periodically shaking potentials used to trap ultra-cold atoms is currently established as an
experimental method~\cite{LignierEtAl07,SiasEtAl08,KierigEtAl08}: 
tunneling has been suppressed by choosing the correct amplitude of driving both for a
Bose-Einstein condensate (BEC) in an optical lattice~\cite{LignierEtAl07,EckardtEtAl05,EckardtEtAl09b} and for a single
atom in a double well potential~\cite{GrifoniHanggi98,KierigEtAl08}. By fine-tuning the driving
frequency, tunneling in tilted systems can also be enhanced in an analogue of
photon-assisted tunneling. After the driving frequency has helped to bridge potential differences,
tunneling can again be controlled by tuning the driving amplitude~\cite{SiasEtAl08,EckardtEtAl05,CreffieldMonteiro06}. Bose-Einstein condensates in periodically driven systems have been investigated regarding tunneling~\cite{JinasunderaEtAl06,StrzysEtAl08}, dynamic localization~\cite{EckardtEtAl09b} and stabilities of exact Floquet states~\cite{HaiEtAl08}.

Suggestions how mesoscopic superpositions like the NOON-state~\cite{Wildfeuer07} 
\begin{equation}
\label{eq:NOON}
|\Psi\rangle\equiv \frac1{\sqrt2}\left(|N,0\rangle+|0,N\rangle\right)
\end{equation}
could be obtained with Bose-Einstein condensates can be found in Refs.~\cite{CastinDalibard97,RuostekoskiEtAl98,CiracEtAl98,DunninghamBurnett01,MicheliEtAl03,Bach04,MahmudEtAl03,WeissTeichmann07,Creffield2007,Dounas-frazerEtAl07,Ferrini2008,WeissCastin09,streltsov-2008,StreltsovEtAl09,DagninoEtAl09}; for a Bose-Einstein condensate in a double well this corresponds to a quantum superposition of all $N$ particles being either in one well or in the other. Here $|n,N-n\rangle$ denotes the Fock state with $n$ particles in the left well and $N-n$ particles in the right well. Like the spin-squeezed states investigated in Refs.~\cite{SorensenEtAl01,EsteveEtAl08,LiEtAl08} such states are relevant to improve interferometric measurements~\cite{GiovannettiEtAl2004}.

A promising approach to obtain mesoscopic superpositions are the
collapse and revival phenomena in phase-space if the tunneling between both wells of a double-well potential is suppressed (or alternatively, for a single-species condensate in a harmonic trapping potential the transition between two hyperfine-states). A NOON-state then appears on the time scale~\cite{HarocheRaimond06}
\begin{equation}
t_{0}=\frac{\pi}{4\kappa}
\end{equation}
where $2\hbar\kappa$ is the interaction energy of a pair of atoms.
Chaos-induced entanglement generation in a periodically shaken double well should, in principle, offer the possibility to obtain mesoscopic superpositions on short time scales~\nocite{WeissTeichmann08}\cite{WeissTeichmann08}. While the state will not be a perfect superposition, the time scale on which it is reached will be an order of magnitude smaller. In this paper, the question is if these superpositions can be achieved both for experimentally realistic conditions and for parameters such that the superposition is reproducible: in an experiment the particle number will vary from shot to shot and the tilt in a double well or the amplitude of shaking is likely to be slightly different.

The paper is organized as follows: Section~\ref{sec:a} introduces the model used to describe a Bose-Einstein condensate in a periodically shaken double well. Section~\ref{sec:char} describes experimental signatures of the entangled states. In Sec.~\ref{sec:emer} a generator of mesoscopic superpositions is discussed while Sec.~\ref{sec:robu} demonstrates that the generation of mesoscopic quantum superpositions is robust against slight changes of the initial conditions.

\section{\label{sec:a}A BEC in a periodically shaken double well}
Bose-Einstein condensates in double-well potentials are interesting both
experimentally and
theoretically~\cite{AlbiezEtAl05,SmerziEtAl97,CastinDalibard97,LesanovskyEtAl06,PiazzaEtAl2008,LeeEtAl08,EsteveEtAl08}. In order to describe a BEC in a double well, we use a model
originally developed in nuclear physics~\cite{LipkinEtAl65}: a many-particle Hamiltonian in two-mode
approximation~\cite{MilburnEtAl97}, 
\begin{eqnarray}
\label{eq:H}
\hat{H} &=& -\frac{\hbar\Omega}2\left(\hat{c}_1^{\dag}\hat{c}_2^{\phantom\dag} +\hat{c}_2^{\dag}\hat{c}_1^{\phantom\dag}\right) + \hbar\kappa\left(\hat{c}_1^{\dag}\hat{c}_1^{\dag}\hat{c}_1^{\phantom\dag}\hat{c}_1^{\phantom\dag}+\hat{c}_2^{\dag}\hat{c}_2^{\dag}\hat{c}_2^{\phantom\dag}\hat{c}_2^{\phantom\dag}\right)\nonumber\\
&+&\hbar\big(\mu_0+\mu_1\sin(\omega t)\big)\left(\hat{c}_2^{\dag}\hat{c}_2^{\phantom\dag}-\hat{c}_1^{\dag}\hat{c}_1^{\phantom\dag}\right)\;,
\end{eqnarray}
where the operator $\hat{c}^{(\dag)}_j$  annihilates (creates) a boson in well~$j$;
$\hbar\Omega$ is the tunneling splitting, $2\hbar\mu_0$ is
the tilt between well~1 and well~2 and $\hbar\mu_1$ is the driving amplitude. The interaction energy of a pair of particles in the same well is denoted by $2\hbar\kappa$.

The Gross-Pitaevskii dynamics can be mapped to that of a nonrigid
pendulum~\cite{SmerziEtAl97}. Including the term  describing the  periodic shaking, the 
classical Hamiltonian is given by:
\begin{eqnarray}
\label{eq:mean}
H_{\rm mf}& = &\frac{N\kappa}{\Omega}z^2-\sqrt{1-z^2}\cos(\phi)\nonumber\\
&-&2z\left(\frac{\mu_0}{\Omega}+
\frac{\mu_1}{\Omega}\sin\left({\textstyle\frac{\omega}{\Omega}}\tau\right)\right)\;,\quad \tau =t\Omega\;,
\end{eqnarray}
where $z$ is the population imbalance with $z=1$ \mbox{($z=-1$)} referring to the situation with all
particles in well~1 (well~2).

 On the $N$-particle quantum level,
if all atoms occupy the single-particle state characterized by $z=\cos(\theta)$ and $\phi$, this  leads to the wave-function
\begin{eqnarray}
\label{eq:ACS}
\left|\theta,\phi\right>&=& \sum_{n=0}^N 
\left({N}\atop{n}\right)^{1/2}
\cos^{n}(\theta/2)
                 \sin^{N-n}(\theta/2)
                 \nonumber\\
                 &\times& e^{i(N-n)\phi}| n, N-n \rangle\;.
\end{eqnarray}
These bimodal phase-states~\cite{HarocheRaimond06} are sometimes referred to as atomic coherent states. Note that for finite $N$ these are not orthogonal:
\begin{equation}\label{eq:ACS2}
|\langle \theta^{\phantom{'}},\phi^{\phantom{'}} | \theta',\phi' \rangle |^2 > 0,\ N<\infty.
\end{equation}

As the Gross-Pitaevskii dynamics corresponds to that of a non-rigid pendulum, the dynamics is known to display a coexistence between chaotic and regular regions~\cite{GuckenheimerHolmes83}. Here, the relation to entanglement~\cite{WeissTeichmann08} is discussed (cf.\ \cite{GhoseEtAl08,TrailEtAl08} for the delta-kicked top).

\section{\label{sec:char}Signatures of mesoscopic quantum superpositions}

The signatures of entangled states described in this section can only serve as signatures of entangled states~\cite{WeissTeichmann09} if they are followed by a revival. However, as these revivals are only partial, this will not be a proof of entanglement but nevertheless a strong indication. Furthermore, on the level of computer simulations, one does not have the necessity to distinguish between quantum superpositions and statistical mixtures. Deepening information concerning the states of interest, superpositions of atomic coherent states~\eqref{eq:ACS}, can be found in \cite{WeissTeichmann08}.

One signature of entanglement is the quantum Fisher information~\cite{PezzeSmerzi09,footnoteQFIneu3}, here for the relative phase between the condensates in the two potential wells (in the following abbreviatory referred to as the quantum Fisher information). Defining for pure states
\begin{equation}
\label{eq:QFI}
 F_{\rm QFI}\equiv {\left(\Delta n_{12}\right)^2}\;,
\end{equation}
where $\left(\Delta n_{12}\right)^2$ is the experimentally measurable~\cite{EsteveEtAl08}
variance of the particle number difference between both wells, one has an entanglement
flag. For pure states,
\begin{equation}\label{eq:QFI2}
 F_{\rm ent} >1,\,\quad F_{\rm ent}\equiv  \frac{F_{\rm QFI}}N
\end{equation}
is a sufficient condition for particle-entanglement and identifies those entangled states that are useful to overcome classical phase sensitivity in interferometric measurements~\cite{PezzeSmerzi09}. Defining entangled states as states which cannot be written as product states~\cite{HarocheRaimond06}, this statement is
supported by the fact that for a single atomic coherent state~(\ref{eq:ACS}) one has $F_{\rm
  ent}\le 1$. This is expected since in this case all atoms occupy the same single-particle state. For mixtures, the quantum Fisher information becomes more complicated (Ref.~\cite{PezzeSmerzi09} and references therein) and might not be easy to measure. Taking the definition~(\ref{eq:QFI}) for an experiment leaves the difficulty to distinguish superpositions from mixtures.

Another experimental signature~\cite{footnoteCoM} will be provided by interference 
patterns after switching off the potential and letting the wave-function expand for some time. A single atomic
coherent state $|\theta,\phi\rangle$ leads to an interference pattern used to
experimentally detect the phase between the condensates~\cite{AlbiezEtAl05}
\begin{equation}
\label{eq:I}
I(X)=[1+\sin(\theta)\cos(X-\phi)]
\end{equation}
which still has to be multiplied by a Gaussian envelope; $X$ essentially is the spatial
variable in the direction connecting the two wells of the double well and $I$ the intensity at $X$.
To characterize the quality of interference fringes, the visibility or contrast
\begin{equation}
\label{eq:visibility}
C \equiv \frac{I_{\rm max}-I_{\rm min}}{I_{\rm max}+I_{\rm min}}
\end{equation}
is introduced where $I_{\rm max}$ is the highest and  $I_{\rm min}$ the lowest value of $I$. Given the fact
that even a single atomic coherent state can produce an impressive interference pattern,
interference cannot be used in a straightforward manner as an entanglement flag as in
Refs.~\cite{PiazzaEtAl2008,WeissCastin09}. Let us thus consider the \textit{disappearance}\/ of
interference patterns,
\begin{equation}
\label{eq:Iweg}
C\ll 1\;.
\end{equation}
This might simply mean $\sin(\theta)\simeq 0$
(cf.\ Eq.~(\ref{eq:I})) or even heating. However, when combined with a high quantum Fisher information, it will be the
signature for the superposition of two (or more) distinct atomic coherent states
with $\sin(\theta_1)\simeq\sin(\theta_2)$ and $\cos(X-\phi_1)+\cos(X-\phi_2)=0$. This is a strong indication for mesoscopic superpositions similar to the case of the superfluid to Mott-insulator transition \cite{GreinerEtAl02} where the dis- and reappearance of the contrast clearly indicates (but does not proof) the Mott-insulator transition.

In order to numerically
calculate the contrast, one needs~\cite{SinatraCastin00}
\begin{equation}
\langle\psi|\hat{\Psi}^{\dag}(x)\hat{\Psi}(x)|\psi\rangle\;
\end{equation}
with
\begin{equation}
 \hat{\Psi}(x) = \Phi_1(x)\hat{c}_1 + \Phi_2(x)\hat{c}_2,
\end{equation}
where $\Phi_i$ is the expanded mode which was originally localized in well $i$ with $i=1,2$ before switching off
the potential.

\section{\label{sec:emer}Emergence of quantum superpositions}
\begin{figure}
\includegraphics[width=1\columnwidth]{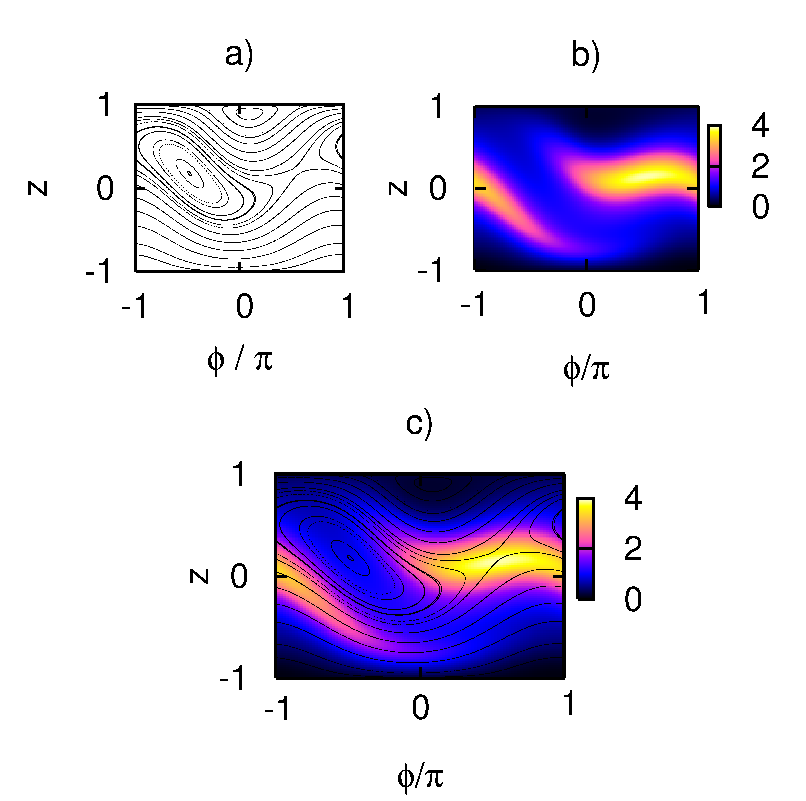}
\caption{\label{fig:niklas}(Color online) Comparison of Poincar\'e surface of section (a) and QFI-map (b) for parameters
$2\mu_0/\Omega=3.0$, $N\kappa/\Omega = 0.5$, $\omega/\Omega = 3.0$ and $2\mu_1/\omega=0.1$.
Plot (b) shows the time averaged QFI for
a system with $N=100$ particles and averaging time $10/\Omega$.
The QFI is enhanced close to the hyperbolic fixed point in the Poincar\'e
section of surface ($z \approx 0.21$, $\phi \approx 0.54\pi$), which acts as a generator of mesoscopic entanglement (c).}
\end{figure}
The claim of Ref.~\cite{WeissTeichmann08} is not that mean-field chaos is the origin of entanglement generation but rather that it speeds up the emergence of mesoscopic superpositions. Thus, rather than looking for entanglement generators which are only present in mean-field chaos, at least some of the aspects should be understandable in a parameter regime for which the mean-field dynamics is regular and thus the corresponding Poincar\'e surface of section (cf.\ Fig.~\ref{fig:niklas}a) is easy to understand: essentially, there is one elliptic fixed point at $z \approx 0.17$, $\phi \approx -0.48\pi$ and one hyperbolic fixed point near $z \approx 0.21$ and $\phi \approx 0.54\pi$, in addition there are further fixed points near the top and bottom of the plot which can be discarded at this point. Within the mean-field dynamics, a solution which starts right on a hyperbolic fixed point ($\theta_{\rm hyper}$, $\phi_{\rm hyper}$) will not move at all. This, however, is different on the $N$-particle quantum level: preparing the system in the atomic coherent state corresponding to all $N$ particles being in the atomic coherent state $|\theta_{\rm hyper},\phi_{\rm hyper}\rangle$ will - for any $N<\infty$ - have contributions from neighboring atomic coherent states, cf. Eq.~\eqref{eq:ACS2}. Those states, even on the mean-field level, do move.

Figure~\ref{fig:niklas}b shows the time-averaged quantum Fisher information of all particles initially occupying an atomic coherent state. For starting values in the vicinity of the hyperbolic fixed point mentioned above, mesoscopic superpositions occur. If this is indeed what is happening, the remaining question is why the other hyperbolic fixed points do not display such a behavior: contrary to the first one, they do not lie near the separatrix for which the dynamics happens on much shorter time scales.

Related to the observed acceleration of entanglement generation is the identification of enhanced atomic tunneling between the two wells in high-chaoticity regions \cite{RongEtAl09}.

\section{\label{sec:robu}Robustness of entanglement generation}
In this section parameters are presented where an entangled state occurs on short time scales, indicated by both signatures quantum Fisher information and contrast. The robustness of this state under realistic experimental conditions is discussed. In the following the initial state is prepared without driving, hence corresponds to the thermal occupation of the eigenstates with energies $E_i$ of Hamiltonian \eqref{eq:H} exclusive of the last term. In a typical experimental situation only some energy eigenstates are significantly occupied~\cite{EsteveEtAl08} which could considerably be reduced by working at lower temperatures in the sub nano Kelvin range~\cite{LeanhardtEtAl03}. The initial state can be prepared in a well-controlled manner and the experimental parameters are well controllable~\cite{EsteveEtAl08} except for small experimental uncertainties. The occurring fluctuations of the initial particle number are discussed in this section while particle losses are not analyzed; their effect on phase revivals has been investigated in \cite{SinatraCastin98}.

Figure~\ref{fig:maps_Fent_C} shows the time development of quantum Fisher information and contrast for initially ground and first excited state in dependence of the scaled driving amplitude $2\mu_1/\omega$. Obviously both initial states show a similar behavior. Parameters where both quantum Fisher information and contrast strongly indicate entanglement are shown in Fig.~\ref{fig:cut_Fent_C}. The quantum Fisher information takes on a value of 70.5 and the contrast decreases to 1~\% already at $\tau=7.1$. Additionally the contrast reappears a short time later, hence heating can be excluded in an experiment.

{\begin{figure}
	\includegraphics[width=1\columnwidth]{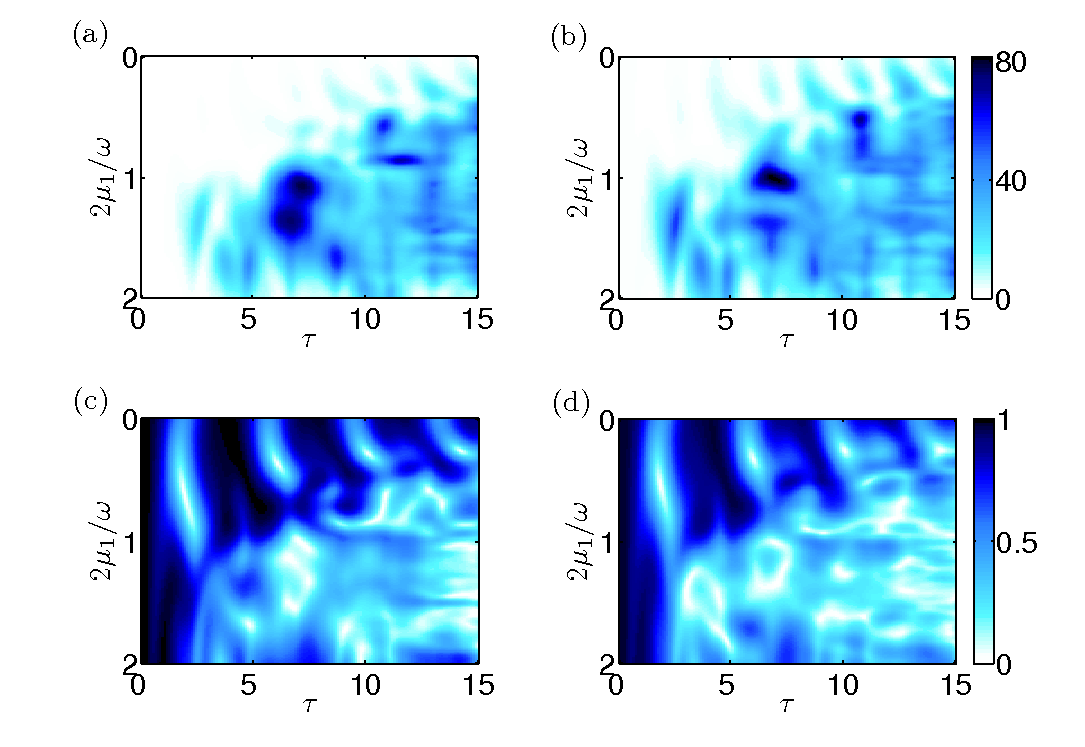}
	\caption{(Color online) Entanglement flag $F_{\mathrm{ent}}$ as a function of time $\tau = \Omega t$ and scaled driving amplitude $2\mu_1/\omega$ for ground (a) and first excited state (b). (c) and (d): Contrast for ground and first excited state. \mbox{$N$ = 110}, \mbox{$N\kappa/\Omega$ = 1.045}, \mbox{$\mu_0/\Omega$ = 0.75}, \mbox{$\omega/\Omega$ = 3.0.}}
	\label{fig:maps_Fent_C}
\end{figure}}
{\begin{figure}
	\includegraphics[width=\columnwidth]{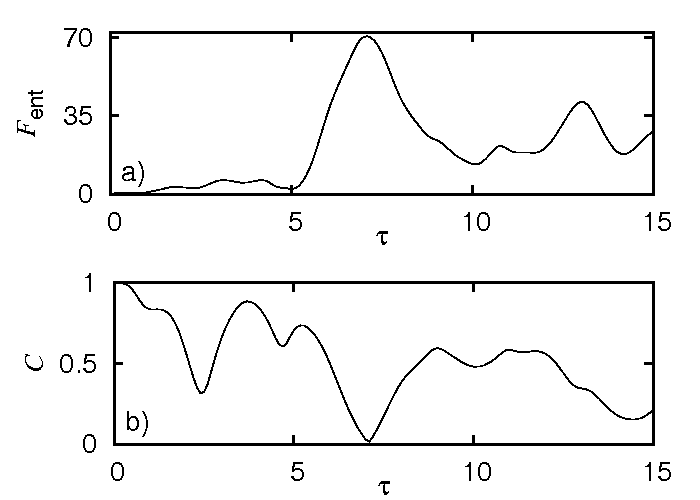}
	\caption{Entanglement flag $F_{\mathrm{ent}}$(a) and contrast (b) for \mbox{$2\mu_1/\omega$ = 1.0}. Other parameters as in Fig.~\ref{fig:maps_Fent_C}.}.
	\label{fig:cut_Fent_C}
\end{figure}}
To determine quantum Fisher information or contrast in an experiment it has to be repeated several times with the same initial conditions. But in the laboratory parameters like driving frequency $\omega$ and tilt $2\hbar\mu_0$ will change slightly from shot to shot. Figure~\ref{fig:mapmu0omega} shows the influence of slight deviations on the two entanglement indicators. Another parameter where deviations will occur is the initial particle number $N$. The dependence of the signatures is represented in Figs.~\ref{fig:varN}. For Poissonian distributed initial particle numbers the signatures are likely to be still visible in the experiment.

So far, we have optimized with respect to interference patterns and fluctuations of the initial particle number $N$. Other experimentally interesting quantum states are characterized by a bimodal phase-distribution with a small standard deviation. Using the completeness relation \cite{MandelWolf95}
\begin{equation}
\mathbf{1} = \frac{N+1}{4\pi}\int \mathrm{d}\theta \sin(\theta) \int \mathrm{d}\phi |\theta,\phi\rangle\langle\theta,\phi|
\end{equation}
the phase distribution $p_{\phi}$ of a state can be calculated. The initial particle number is again assumed to obey a Poissonian distribution with mean value $N = 110$ which for numerical purposes has been truncated and renormalized with $N$ ranging from $N = 100$ to $N = 120$. Then, a mean phase distribution and its standard deviation can be calculated for each point of time which has been optimized in order to find a phase distribution that indicates strong entanglement and as well has a small standard deviation. In such an optimization procedure an entangled state has been found at $\tau =6.5$ where the phase distribution shows two distinct maxima, separated by $\Delta \phi \approx \pi$, cf. Fig.~\ref{fig:phase}b. For maximum reappearance one important peak is observed at $\tau = 8.9$ (c). For comparison, an exemplary mixed state leading to the phase distribution of (b) has been constructed: the wavefunction at $\tau =6.5$ has been split into two parts, each consisting of the contributions to $p_{\phi}$ in an interval of $\Delta \phi = \pi$ around one of the peaks. In the statistical mixture these two, renormalized parts are weighted according to their contribution to the total wavefunction. This mixed state would after the corresponding time difference $\Delta \tau = 2.4$ have a phase distribution according to (d) which is distinct of (c). Hence, it is possible to distinguish pure entangled states from mixed states. Additionally, preliminary results indicate that under slight parameter variations of about 1 \% for driving frequency, driving strength and initial tilt respectively 10 \% for the scaled interaction strength $N\kappa/\Omega$, the phase distribution at $\tau = 6.5$ and the partial revival at $\tau = 8.9$ stay nonetheless robust. 

{\begin{figure}
	\includegraphics[width=\columnwidth]{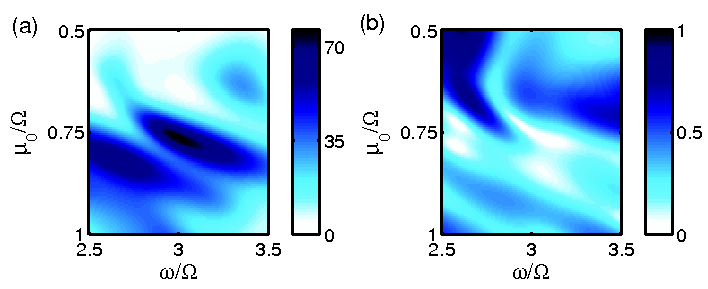}
	\caption{(Color online) Entanglement flag (a) and contrast (b) at \mbox{$\tau$ = 7.1} versus $\mu_0/\Omega$ and $\omega/\Omega$. \mbox{$N$ = 110}, \mbox{$N\kappa/\Omega$ = 1.045}, \mbox{$2\mu_1/\omega$ = 1.0}.}
	\label{fig:mapmu0omega}
\end{figure}}
{\begin{figure}
	\includegraphics[width=\columnwidth]{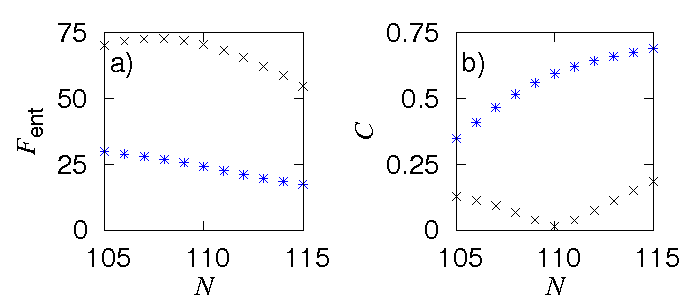}
	\caption{(Color online) $F_{\textrm{ent}}$ (a) and contrast (b) for \mbox{$\tau$ = 7.1} (black crosses) and \mbox{$\tau$ = 9.0} (blue stars) for different particle numbers such that $\kappa/\Omega$ = const. \mbox{$N\kappa/\Omega (N = 110)$ = 1.045}, \mbox{$\mu_0/\Omega$ = 0.75}, \mbox{$\omega/\Omega$ = 3.0}, \mbox{$2\mu_1$ = 1.0}.}
	\label{fig:varN}
\end{figure}}
{\begin{figure}
	\includegraphics[width=\columnwidth]{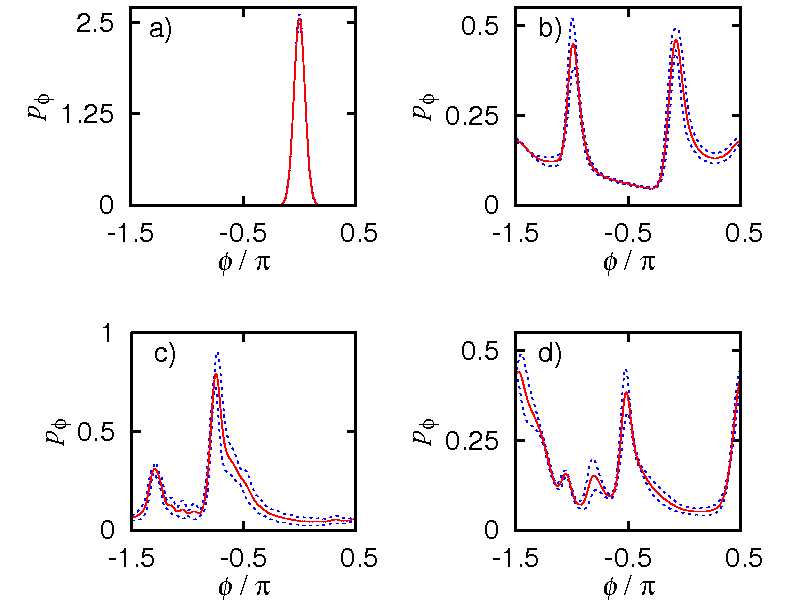}
	\caption{(Color online) Averaged phase distribution $p_{\phi}$ (red solid line) with standard deviation (blue dashed lines). The initial particle number is modeled by a truncated, renormalized Poisson distribution of the initial particle number, ranging from $N =$ 100 to $N=$ 120, with mean value \mbox{$N=$ 110.} \mbox{(a) Initial state,} (b) entangled state at $\tau = 6.5$, (c) time-developed state at $\tau = 8.9$, (d) assumed exemplary, two-component statistical mixture at $\tau = 8.9$. The interaction strength $\kappa/\Omega$ is kept constant such that \mbox{$N\kappa/\Omega (N = 110)$ = 1.0}. \mbox{$\mu_0/\Omega$ = 1.3}, \mbox{$\omega/\Omega$ = 3.3}, \mbox{$2\mu_1/\omega$ = 0.7}.}
	\label{fig:phase}
\end{figure}}
{\begin{figure}
	\includegraphics[width=\columnwidth]{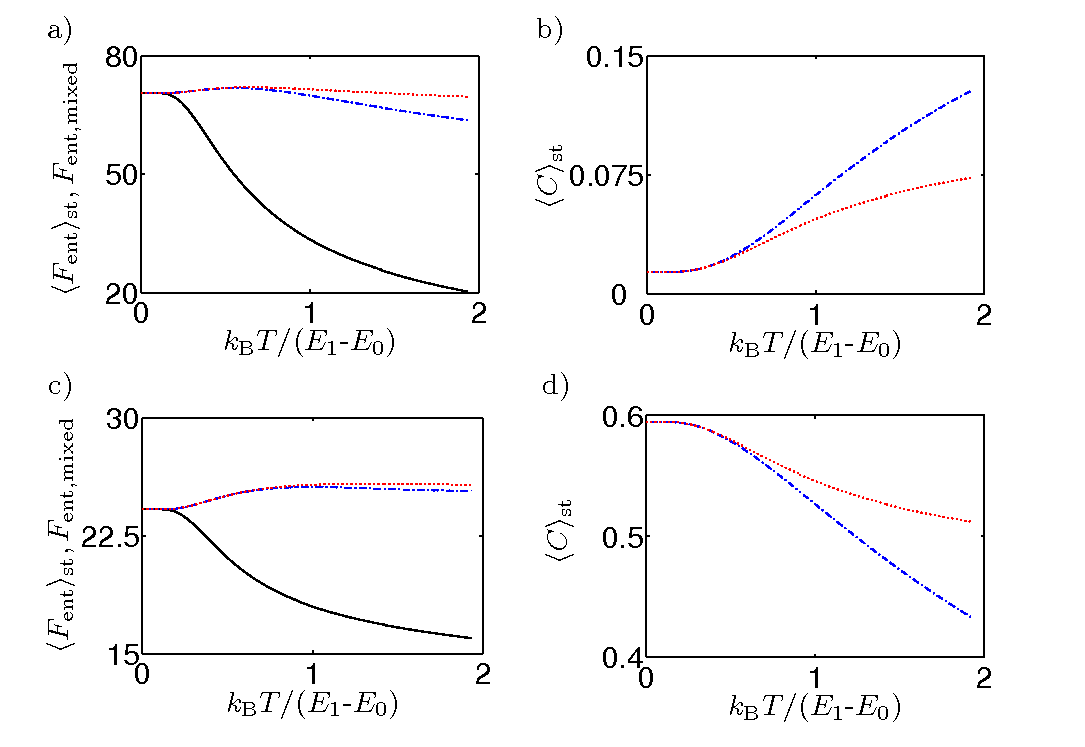}
	\caption{(Color online) $F_{\textrm{ent}}$ for $\tau$ = 7.1 (a) and \mbox{$\tau$ = 9.0 (c)}. Black, solid line: quantum Fisher information $F_{\mathrm{ent,mixed}}$ for mixed states (cf. Eq. \eqref{eq:Fentmixed}), blue, dashed line: $\langle F_{\mathrm{ent}} \rangle_{\mathrm{st}}$ statistically averaged over all 111 states, red, dotted line: $\langle F_{\mathrm{ent}} \rangle_{\mathrm{st}}$ statistically averaged over three lowest lying, experimentally important states \cite{EsteveEtAl08}. (b) and (d): contrast $\langle C\rangle_{\mathrm{st}}$ for \mbox{$\tau$ = 7.1} and \mbox{$\tau$ = 9.0} (cf. Eq. \eqref{eq:Caverage}). Blue, dash-dotted line: statistical average over all states, red, dotted line: statistical average over three lowest lying eigenstates. Parameters same as in Fig.~3, \mbox{$2\mu_1/\omega$ = 1.0}.}
	\label{fig:temperature_effects}
\end{figure}}
Besides slight changes in the experimental conditions another effect that should be discussed is the influence of temperature. So far, calculations have been performed for $T=0$. To account for temperature the eigenstates of the undriven system corresponding to energies $E_i$ are supposed to be populated with probability $p_i$ according to a Boltzmann distribution
\begin{equation}
p_i = \frac{\mathrm{e}^{-(E_i - E_0)/k_\mathrm{B} T}}{\mathcal{Z}}, \ \mathcal{Z} = \sum_{i=0}^M \mathrm{e}^{-(E_i - E_0)/k_\mathrm{B} T}.
\end{equation}
Hence, the statistically averaged contrast is given by 
\begin{equation}\label{eq:Caverage}
\langle C \rangle_{\mathrm{st}} = \sum^M_{i=0} p_i C_i
\end{equation}
where $C_i$ is the contrast for the pure state $i$ and $M$ is the number of states included in the average. With an analogue definition of $\langle \cdot \rangle_{\mathrm{st}}$ the statistically averaged quantum Fisher information becomes
\begin{equation} \label{eq:statavQFI}
\langle F_{\mathrm{ent}} \rangle_{\mathrm{st}} = \frac{\langle \left( \Delta n_{12} \right)^2  \rangle_{\mathrm{st}}}{N} = \frac{1}{N}  \left( \langle n^2_{12}  \rangle_{\mathrm{st}} - \langle n_{12}  \rangle^2_{\mathrm{st}} \right).
\end{equation}
For comparison also the quantum Fisher information for mixed states (Ref~\cite{PezzeSmerzi09} and references therein)
\begin{equation}\label{eq:Fentmixed}
F_{\mathrm{ent,mixed}}=\frac{1}{2N}\sum_{j,k}\frac{\left( p_j - p_k \right)^2}{p_j + p_k} \left| \langle j | n_{12} | k  \rangle  \right|^2,
\end{equation}
again relative to the relative phase, is calculated. Temperature effects are represented in Fig.~\ref{fig:temperature_effects} for the same parameters as in Fig. 2 and the interesting points of time $\tau = 7.1$ and $\tau = 9.0$ where the contrast vanishes respectively reappears maximally. As the experiments in \cite{EsteveEtAl08} are performed in a temperature regime where only the three lowest lying states are significantly populated both the statistical averages over the three lowest lying states and all states are calculated. The curves for only the three lowest lying and all states stay almost identical up to temperatures where $k_{\mathrm{B}}T$ approximately equals the energy separation of ground and first excited state. The value of the quantum Fisher information at $\tau = 7.1$ is still adequate while the contrast stays low. Also for $\tau = 9.0$ the signatures are not influenced too much. For higher temperatures the quantum Fisher information for mixed states deviates from the statistically averaged quantum Fisher information \eqref{eq:statavQFI}. Hence, it should be noted that measuring the variance of the particle number difference at higher temperatures is not equivalent to measuring the quantum Fisher information for mixed states while for low enough temperatures both quantities lead to similar results. 

\section*{Conclusion}
To summarize, the occurrence of entanglement on short time scales is observed for a Bose-Einstein condensate in a tilted double-well potential. For this system on the mean-field level the entanglement generation is accelerated in the vicinity of hyperbolic fixed points. The robustness of the entangled states under realistic experimental conditions is analyzed and parameters are presented where the signatures quantum Fisher information for the relative phase and contrast are found to be sufficiently stable. In conclusion, for the proposed parameters the phase distribution of the entangled state measured in an experiment is expected to exhibit two clearly distinguishable maxima.

Our results are based upon the widely-used two-mode approximation. An analysis of its limitations can be found, e.g., in \cite{SakmannEtAl09}. Nevertheless, the two-mode approximation has, especially for comparatively small interaction strengths, been found to describe many aspects of the experiments \cite{EsteveEtAl08,AlbiezEtAl05}. Besides the investigation of extended models, the influence of particle losses is certainly an interesting aspect for future work.

\acknowledgments
We thank M.~Holthaus and M.~Oberthaler for discussions. 
N.~T. and B.~G. acknowledge funding by the Studienstiftung des deutschen Volkes.



\begin{thebibliography}{53}
\bibitem[{foo({\natexlab{a}})}]{footnoteQFIneu3}
\bibinfo{note}{Here the observable $\vec{J}_{\vec{n}}$ from
  \cite{PezzeSmerzi09} is taken to be the particle number difference $n_{12}$.
  Hence, a quantum Fisher information relative to the relative phase is
  calculated.}
\end{thebibliography}

\end{document}